\documentclass[aps,prb,reprint,superscriptaddress]{revtex4-1}

\usepackage{graphicx}
\usepackage{hyperref}
\usepackage{xcolor}

\begin{document}


\title{Selective tuning of spin-orbital Kondo contributions in parallel-coupled quantum dots}


\author{Heidi Potts}
\email[]{heidi.potts@ftf.lth.se}
\author{Martin Leijnse}
\author{Adam Burke}
\author{Malin Nilsson}
\author{Sebastian Lehmann}
\affiliation{Division of Solid State Physics and NanoLund, Lund University, SE-221 00 Lund, Sweden}
\author{Kimberly A. Dick}
\affiliation{Division of Solid State Physics and NanoLund, Lund University, SE-221 00 Lund, Sweden}
\affiliation{Centre for Analysis and Synthesis, Lund University, SE-221 00 Lund, Sweden}
\author{Claes Thelander}
\email[]{claes.thelander@ftf.lth.se}
\affiliation{Division of Solid State Physics and NanoLund, Lund University, SE-221 00 Lund, Sweden}


\date{\today}

\begin{abstract}
We use co-tunneling spectroscopy to investigate spin-, orbital-, and  spin-orbital Kondo transport in a strongly confined system of InAs double quantum dots (QDs) parallel-coupled to source and drain. In the one-electron transport regime, the higher symmetry spin-orbital Kondo effect manifests at orbital degeneracy and no external magnetic field. We then proceed to show that the individual Kondo contributions can be isolated and studied separately; either by orbital detuning in the case of spin-Kondo transport, or by spin splitting in the case of orbital Kondo transport. By varying the inter-dot tunnel coupling, we show that lifting of the spin degeneracy is key to confirming the presence of an orbital degeneracy, and to detecting a small orbital hybridization gap. Finally, in the two-electron regime, we show that the presence of a spin-triplet ground state results in spin-Kondo transport at zero magnetic field.
\end{abstract}


\maketitle

\section{Introduction}

The Kondo effect is a widely studied many-body phenomenon that has increased the understanding of strongly correlated electron systems. Experimentally it can be investigated using quantum dots (QDs) with highly transparent tunnel barriers, and manifests as a zero-bias conductance resonance. Most studies focus on the spin-1/2 Kondo effect, where an unpaired spin is screened in the absence of a magnetic field \cite{Goldhaber-Gordon1998,Cronenwett1998,Schmid1998}. Two-electron spin states represent another common system for Kondo studies, where resonances arise from a
vanishing singlet-triplet exchange energy \cite{Schmid2000,Kogan2003,Paaske2006}, a magnetic field induced singlet-triplet crossing \cite{Tarucha2000,Pustilnik2000,Sasaki2000,Giuliano2001,Wiel2002,Chorley2012}, or a triplet ground state \cite{Liang2002,Quay2007,Roch2009}. More recently, the orbital Kondo effect, which relies on two degenerate orbitals, has received considerable attention. In particular, by combining  spin and orbital degrees of freedom, the SU(4) Kondo effect can be studied \cite{Krychowski2016}. The presence of such a higher symmetry can be identified by different temperature scaling \cite{Keller2013}, and an expected enhancement in the shot-noise properties \cite{Delattre2009,Ferrier2015,Ferrier2017}.

In single QDs, orbital and spin-orbital Kondo transport has been observed in carbon nanotubes \cite{Jarillo-herrero2005,Makarovski2007a,Makarovski2007b,Delattre2009,Ferrier2015,Ferrier2017}, and silicon based devices \cite{Lansbergen2010,Tettamanzi2012}. However, in such systems, the orbital degeneracy is an inherent material property and the tunability of the orbital alignment is therefore limited. An alternative approach is to fabricate two parallel-coupled QDs for which the orbitals can be tuned independently. Typically, this is realized in two-dimensional electron gases \cite{Wilhelm2001,Holleitner2004,Sasaki2004,Amasha2013,Keller2013,Okazaki2011}, but at the expense of limited Zeeman splitting of spin states. 
\\

In this work, we use parallel-coupled quantum dots in indium-arsenide (InAs) nanowires to investigate the spin- and orbital Kondo effect. We show that each degeneracy here can  be induced and lifted selectively, which makes it an ideal system for studies of higher Kondo symmetries. We start by discussing Kondo resonances in the one-electron regime, and present results on the spin-, orbital-, and combined spin-orbital Kondo effect. The individual contributions to the transport are isolated by electric field-induced orbital detuning, and magnetic field-induced spin splitting. Furthermore, by controlling the inter-dot tunnel coupling, $t$, we demonstrate that the formation of hybridized states inhibits the orbital Kondo effect. If $t$ is small, the energy gap of the avoided crossing can only be resolved if the spin-degeneracy is lifted. This underlines the importance of isolating different Kondo mechanisms when studying their contribution to Kondo effects of higher symmetry. Finally, we investigate the Kondo effect in the two-electron regime when each of the QDs contains one unpaired electron, and a finite tunnel coupling between the QDs gives rise to two-electron states. A Kondo resonance at zero magnetic field is here observed due to a spin-triplet ground state.\\

\section{Experiment}

Our study is based on InAs nanowires where a QD is defined by a thin zinc-blende (ZB) section between two wurtzite (WZ) barriers \cite{Nilsson2016}. Figure \ref{Fig1}(a) shows a schematic representation of the nanowire and a scanning electron microscopy (SEM) image of a representative device. Metal contacts (Ni/Au 25\,nm/75\,nm) are placed on the outer ZB sections as source and drain\cite{Nilsson2017}. Transport measurements are performed in a dilution refrigerator with an electron temperature $T_\mathrm{e}\approx 70$\,mK. If applied, the magnetic field, $B$, is aligned perpendicular to the substrate plane in this study. Previous studies have shown that two sidegates ($V_{\mathrm{L}}$, $V_{\mathrm{R}}$) and a global backgate ($V_{\mathrm{BG}}$) allow to split the QD into two parallel-coupled QDs, and the electron population on the two QDs can be controlled independently. The system is highly tunable, and the small effective electron mass provides large intra-dot orbital separations \cite{Nilsson2017,Nilsson2018,Thelander2018}. In recent works, we have furthermore demonstrated the formation of ring-like states when coupling the QDs in two points, resulting in a vanishing one-electron hybridization energy \cite{Potts2019a}, and a spin-triplet ground state when each QD contains an unpaired electron. 

\section{Results and Discussion}

\begin{figure}
	\includegraphics{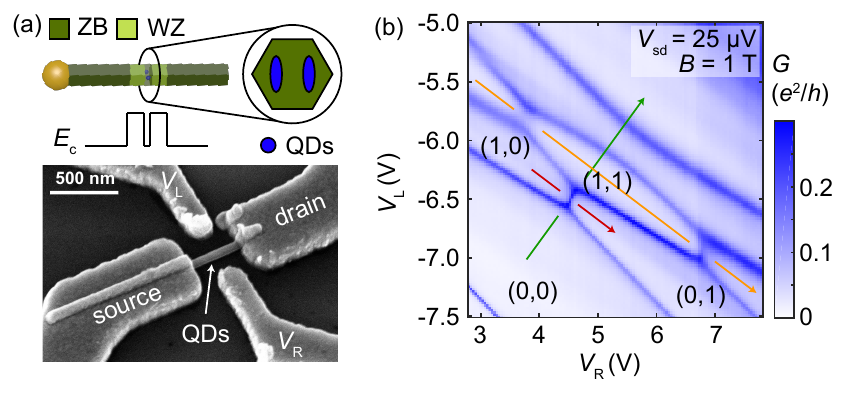}%
	\caption{(a) Top: Schematic of the nanowire and conduction band alignment. Bottom: 45${^\circ}$ tilted SEM image of a representative device. (b) Conductance ($G$) as a function of sidegate voltages at $B$ = 1\,T. The electron populations on the left and right QD ($N_{\mathrm{left}}$,$N_{\mathrm{right}}$) after subtracting 2$N$ electrons are indicated. Important gate vectors are shown in red, green, and orange. \label{Fig1}}
\end{figure}

In this article, we investigate transport in a regime where two orbitals (one from each QD) cross in energy and interact, as shown in Fig. \ref{Fig1}(b). The electron population of the left and right dot ($N_{\mathrm{left}}$,$N_{\mathrm{right}}$) is indicated. Here, 2$N$ electrons were subtracted on both QDs, as the contribution of filled orbitals can be neglected due to strong confinement (a large-range overview measurement can be found in the supporting information). In such an orbital crossing, Kondo transport due to both spin and orbital degeneracies is possible. The spin-1/2 Kondo effect is expected when one of the QDs contains an unpaired electron, which is the case for transport involving the (1,0) and (0,1) orbitals. Assuming a vanishing hybridization energy for these two orbitals, we additionally expect an orbital Kondo effect where they cross in energy. Additionally, in the (1,1) regime, two-electron spin states provide degeneracies that can also result in Kondo transport. In order to study these different types of Kondo origins, we will focus on transport along the gate vectors indicated in green, red, and orange in Fig. \ref{Fig1}(b).\\

\subsection{Spin-orbital Kondo effect}

\begin{figure*}
	\includegraphics{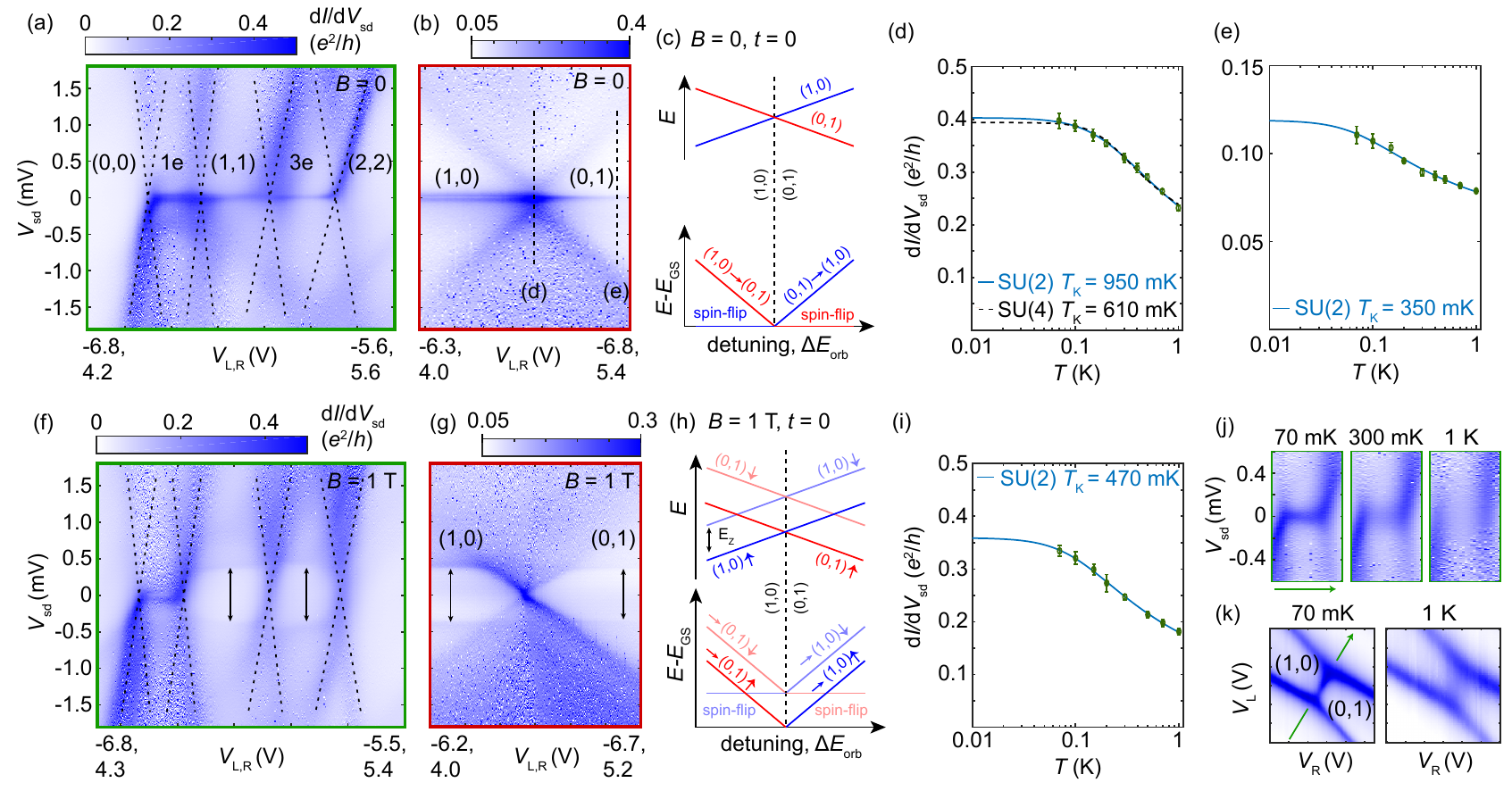}%
	\caption{Spin-orbital and isolated orbital Kondo transport in the one-electron regime. (a) Measurement of d$I$/d$V_\mathrm{sd}$ versus $V_\mathrm{sd}$ along the green gate vector indicated in Fig. \ref{Fig1}(b) for $B = 0$. A zero-bias peak can be observed in the 1e, 2e, and 3e regimes. (b) Corresponding measurement in the 1e regime, when detuning the orbitals along the red gate vector. The dashed lines represent the sidegate voltages where the temperature sweeps are performed. (c) Schematic representation of the electron energy levels and resulting onset of co-tunneling when detuning the QD orbitals (assuming $t=0$). (d) Temperature dependence of the zero-bias conductance peak at orbital degeneracy. Blue and dashed black lines show the fit using Eq. \ref{eq_TK_temp} and the parameters for SU(2) and SU(4) Kondo scaling, respectively. (e) Temperature dependence and SU(2) Kondo fit of the zero-bias conductance peak in the (0,1) state. (f-i) Corresponding measurements and schematic for $B = 1\,T$. The spin degeneracy is lifted, but the orbital degeneracy at zero detuning in the 1e regime remains. (j) Same as (f) but for different temperatures (only 1e regime). (k) Conductance as a function of sidegate voltages ($V_\mathrm{sd}=25$\,$\mu$V) for 70\,mK and 1\,K (1e regime). \label{Fig2}}
\end{figure*}

Figure \ref{Fig2}(a) shows a measurement of differential conductance, d$I$/d$V_{\mathrm{sd}}$, versus the source-drain voltage, $V_\mathrm{sd}$, along the green gate vector indicated in Fig. \ref{Fig1}(b). The gate vector is chosen such that the (1,0) and (0,1) orbitals are approximately degenerate along this vector in the one-electron regime. A zero-bias peak in the differential conductance can be observed within the outlined Coulomb diamonds corresponding to the one-electron (1e), two-electron (2e), and three-electron (3e) regimes. Next, we investigate the 1e regime in more detail by detuning the orbitals of the left and right QD with respect to each other (red gate vector). Due to the large interdot Coulomb energy $U_\mathrm{1,2}\approx 3$\,meV, the system holds one electron along the whole vector, and all transport features are related to co-tunneling events. The transport measurement at $B=0$\,T in Fig. \ref{Fig2}(b) shows a zero-bias peak\cite{note2}, as well as a step in differential conductance that approaches zero bias at zero detuning. Comparing with a schematic representation of the states and resulting onset of co-tunneling transport (Fig. \ref{Fig2}(c)), we can distinguish two transport processes: 1) the spin-1/2 Kondo effect on the populated QD, and 2) co-tunneling involving both QDs, which we will refer to as orbital co-tunneling. For the latter, an electron tunnels out of the left (right) QD and is replaced by an electron on the right (left) QD (with or without a spin-flip), and the energy cost of this process corresponds to the detuning of the two QDs. At zero detuning of the QDs ($\Delta E_\mathrm{orb}=0$), the onset of orbital co-tunneling crosses zero bias. However without further investigation it is unclear whether both the spin and the orbital degeneracy contribute to the Kondo resonance. \\

Transport measurements at different temperatures were performed in order to extract the Kondo temperatures in different locations in the honeycomb diagram. The intensity of a Kondo zero-bias peak shows a characteristic decay with temperature, which can be described by the phenomenological expression \cite{Costi1994, Goldhaber-Gordon1998}
\begin{equation}
G(T)=G_0\left[1+(2^{1/s}-1)\left(\frac{T}{T_\mathrm{K}}\right)^n\right]^{-s}+G_1
\label{eq_TK_temp}
\end{equation}
where $T_\mathrm{K}$ is the Kondo temperature, $G_0$ is the Kondo conductance at $T=0$\,K, and $G_1$ corresponds to a constant background conductance. The parameters $s$ and $n$ depend on the symmetry of the Kondo state. For the spin-1/2 Kondo effect $s=0.22$ and $n=2$ is commonly used (SU(2) parameters), while Keller \textit{et al.} have introduced $s=0.2$ and $n=3$ for the spin-orbital Kondo effect with SU(4) symmetry \cite{Keller2013}. \\

Figure \ref{Fig2}(d) shows the temperature dependence of the conductance at $V_\mathrm{sd}=0$ and $\Delta E_\mathrm{orb}=0$ in the 1e regime. Assuming that both the spin and orbital degeneracies contribute to the Kondo resonance, we use the SU(4) fitting parameters, and extract $T_\mathrm{K}=610$\,mK ($G_0=0.25$\,$e^2$/$h$, $G_1=0.15$\,$e^2$/$h$) using Eq. \ref{eq_TK_temp}. However, also the standard parameters for the SU(2) Kondo effect provide a good fit, and would give a Kondo temperature of $T_\mathrm{K}=950$\,mK ($G_0=0.33$\,$e^2$/$h$, $G_1=0.08$\,$e^2$/$h$), thus making it impossible to conclude the symmetry of the Kondo effect by the temperature dependence alone.

\subsection{Isolating the spin-1/2 Kondo effect}

Next, we isolate spin-Kondo transport in the right QD by electrostatically detuning the system to the (0,1) regime, and study its temperature dependence (Fig. \ref{Fig2}(e)). For the chosen sidegate voltages, $T_\mathrm{K}=350$\,mK can be extracted using Eq. \ref{eq_TK_temp} and standard SU(2) parameters ($G_0=0.06$\,$e^2$/$h$, $G_1=0.06$\,$e^2$/$h$). The fact that $G_0\ll2$\,$e^2$/$h$ can be explained by an asymmetry in the tunnel couplings to source and drain (Appendix \ref{App:Gamma}). In our QD system, the asymmetry can be due to a difference in barrier thickness and shape of the QDs.

\subsection{Isolating the orbital Kondo effect}

In order to unambiguously verify the orbital contribution to the Kondo effect at $\Delta E_\mathrm{orb}=0$, we isolate the effect from spin-Kondo transport using Zeeman spin splitting at $B=1$\,T. The large effective $g$-factor ($g^*$) of InAs (-14.7 in bulk) facilitates lifting of the spin degeneracy by the Zeeman energy $E_\mathrm{Z}=g^*\mu_\mathrm{B}B$, where $\mu_\mathrm{B}$ is the Bohr magneton. In Fig. 2(f) a resulting gap is observed in the 2e and 3e regimes, where spin-flip transport by inelastic co-tunneling is possible when the source-drain voltage is larger than the gap energy. Preliminarily, the absence of zero-bias peaks in the 2e and 3e regimes indicates that the observed Kondo peaks in Figs. \ref{Fig2}(a) were due to a spin-related Kondo effect. However, in the 1e region the zero-bias peak remains, indicating that a different degeneracy is still present\cite{note1}.

Detuning the orbitals along the red gate vector allows to distinguish the Zeeman gap of the single QD orbitals from the onset of orbital co-tunneling (Figs. \ref{Fig2}(g,h)). In the (1,0) and (0,1) regimes, we observe $E_\mathrm{Z}\approx 0.5$\,meV, corresponding to $g^* \approx 9$. The step in differential conductance, originating from the onset of orbital co-tunneling transport, crosses in the Zeeman gap at $\Delta E_\mathrm{orb}=0$. A zero-bias peak can be observed, which corresponds to the orbital Kondo effect. Temperature dependent conductance data at this point (Fig. \ref{Fig2}(i)) can be fitted with Eq. \ref{eq_TK_temp} and standard SU(2) parameters, resulting in $T_\mathrm{K}=470$\,mK ($G_0=0.28$\,$e^2$/$h$, $G_1=0.08$\,$e^2$/$h$). Comparing the Kondo temperature for the spin-orbital Kondo resonance with that of the pure spin-1/2 or the pure orbital Kondo effect, we find a higher $T_\mathrm{K}$ and $G_0$ when both degeneracies are present, which is in agreement with a higher symmetry. 

In Figs. \ref{Fig2}(j-k), measurements at different temperatures are presented for $B=1$\,T. The bias-dependent measurements of d$I$/d$V_\mathrm{sd}$ along the green gate vector in the 1e regime (Fig. \ref{Fig2}(j)) show that the zero-bias peak disappears with increasing temperature, while transport corresponding to sequential tunneling only exhibits a weak temperature dependence. Accordingly, the conductance line at orbital degeneracy strongly decays with temperature in Fig. \ref{Fig2}(k). 

We note that the presence of the orbital Kondo effect implies that the orbital quantum number is also a good quantum number in the leads (similar to spin). In this work, the results are discussed in the picture of coupled double QDs, where the orbital degeneracy is commonly attributed to zero tunnel coupling between the two QDs. An orbital Kondo resonance therefore requires two transport channels in the leads, each of them coupling predominantly only to one of the QDs. However, an orbital degeneracy would also be present if the two QDs are coupled into ring-like states (c.f. \cite{Potts2019a}), if the hybridization gap due to spin-orbit interaction and backscattering is smaller than the Kondo temperature \cite{Jarillo-herrero2005,Delattre2009,Paaske2010}.\\

\begin{figure}
	\includegraphics[width=\columnwidth]{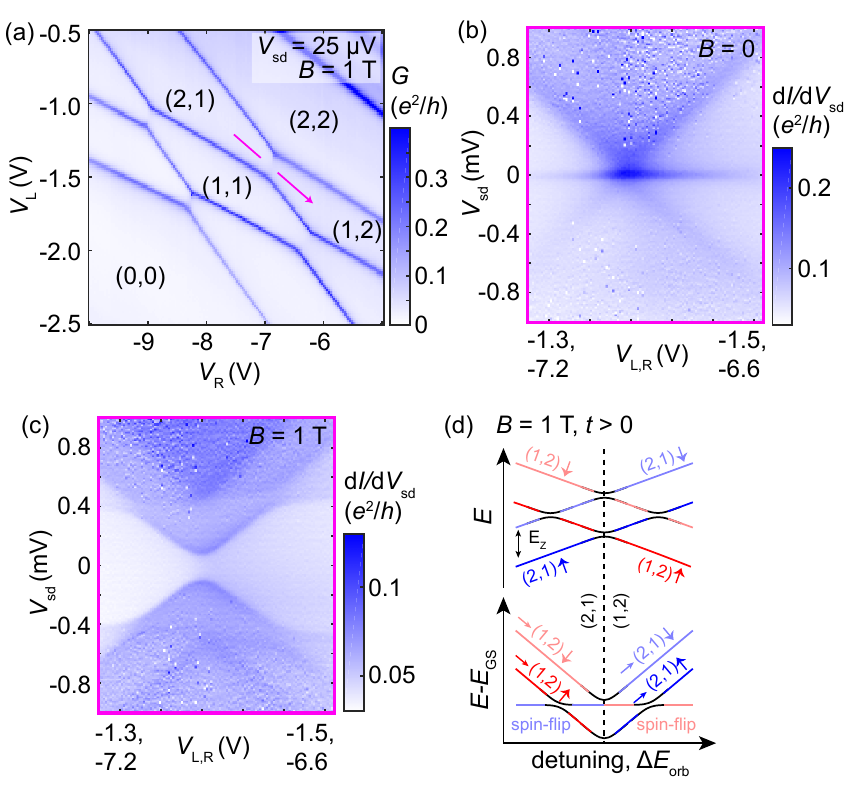}%
	\caption{The effect of a finite tunnel coupling between the QDs in another orbital crossing. (a) Overview measurement with the electron number on the left and right QD ($N_\mathrm{left}$,$N_\mathrm{right}$) indicated. (b,c)  Measurement of d$I$/d$V_\mathrm{sd}$ versus $V_\mathrm{sd}$ recorded along the pink gate vector at $B=0$\,T, and  $B=1$\,T. (d) Schematic representation of the energy levels and resulting onset of co-tunneling when detuning the QD orbitals, assuming finite tunnel coupling between the two QD levels. \label{Fig3}}
\end{figure}

\subsection{Effect of orbital hybridization}

In the following, we will discuss the effect of inter-dot tunnel coupling, based on data from the same device but from a different crossing of QD orbitals (Fig. \ref{Fig3}(a)). When detuning the orbitals along the pink gate vector for $B=0$, both a zero-bias peak, and a sloped step in differential conductance due to orbital co-tunneling can be observed (Fig. \ref{Fig3}(b)). From this data it appears as if the onset of orbital co-tunneling crosses at zero bias for $\Delta E_\mathrm{orb}=0$, similar to what has been shown in Fig. \ref{Fig2}(b).

However, when the spin degeneracy is lifted ($B=1$\,T, Fig. \ref{Fig3}(c)), a finite gap at zero detuning is visible instead of an orbital Kondo resonance. This can be understood considering an avoided crossing due to inter-dot tunnel coupling, as schematically presented in Fig. \ref{Fig3}(d). In this case the hybridization gap is $\sim70\,\mu$eV, corresponding to a temperature of $\sim800$\,mK. Since the gap is comparable to the observed Kondo temperatures, it can only be detected by first lifting the spin-degeneracy. A similar behavior but with a much larger avoided crossing is found for the 3e regime of Fig. \ref{Fig1}(b) (see Appendix Fig. \ref{FigSI1}). \\

\subsection{Spin-1 Kondo transport}

\begin{figure}
	\includegraphics{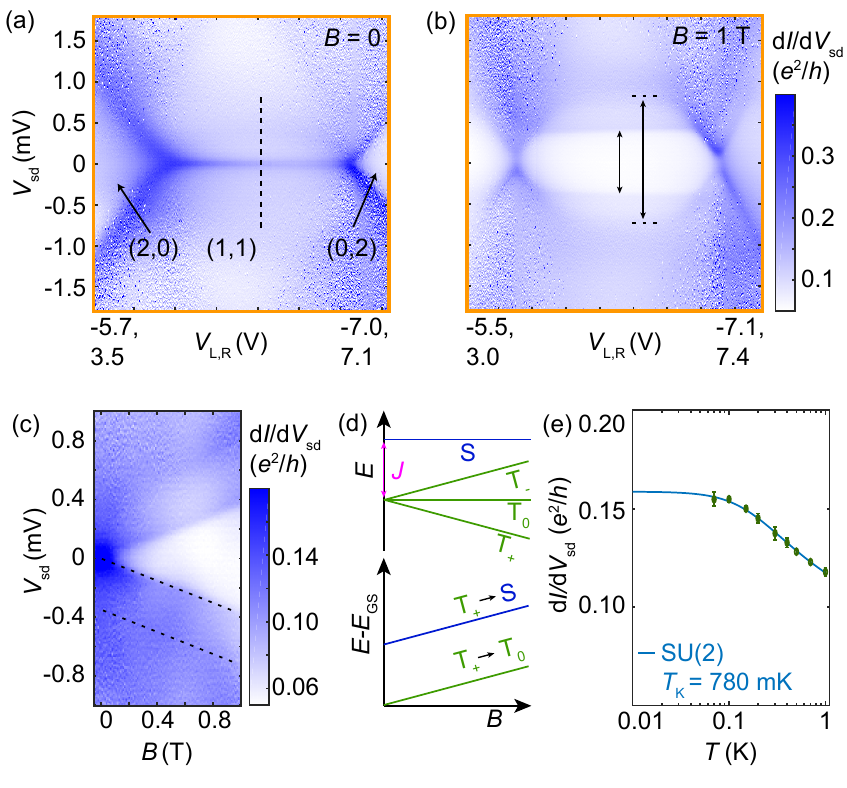}%
	\caption{Kondo transport in the two-electron regime. (a) Measurement of d$I$/d$V_\mathrm{sd}$ versus $V_\mathrm{sd}$ recorded along the orange gate vector at $B=0$\,T, showing a Kondo resonance in (1,1). (b) Corresponding measurement at  $B=1$\,T. The ground state degeneracy is lifted, and an additional onset of co-tunneling is observed with an energy of $\sim350\,\mu$eV higher than the first excited state. (c) Magnetic field sweep in the center of (1,1). (d) Schematic representation of the 2e state energies and resulting onset of co-tunneling as a function of $B$-field. The triplet GS Zeeman splits into T$_+$, T$_0$ and T$_-$. At finite $B$-field, T$_+$ is the GS, and we observe transitions to T$_0$ and to the singlet state, S. (e) Temperature dependence of the zero-bias conductance peak in the center of (1,1) at $B=0$\,T. \label{Fig4}}
\end{figure}

Finally, we investigate Kondo transport in the 2e regime by detuning the orbitals along the orange gate vector in Fig. \ref{Fig1}(b). A zero-bias peak can be observed in the (1,1) regime at $B=0$\,T (Fig. \ref{Fig4}(a)), indicating a Kondo resonance due to a degenerate ground state when each QD contains one electron. At $B=1$\,T (Fig. \ref{Fig4}(b)) this degeneracy is lifted, resulting in a gap of $\sim800\,\mu$eV in the (1,1) regime. A second excited state can be observed $\sim350\,\mu$eV higher in energy than the first excited state,  which implies the existence of two-electron states due to hybridization, rather than non-interacting orbitals on each QD. To study the two-electron states in more detail, bias-dependent transport as function of magnetic field was performed in the center of (1,1), as shown in Fig. \ref{Fig4}(c). The zero-bias peak Zeeman splits with $g^*\approx7$, which is slightly different compared to $g^*\approx9$ in the 1e regime, likely due to the interaction with another orbital \cite{Csonka2008}. The second onset of co-tunneling evolves parallel to the first excited state. These observations are in agreement with a spin-triplet ground state and a spin-singlet excited state, with an exchange energy of $J\approx350\,\mu$eV, as schematically depicted in Fig. \ref{Fig4}(d). The spin-triplet ground state could be explained by the formation of a quantum ring \cite{Pfannkuche2014}, which can form when two quantum dots couple in two points, as previously shown for this material system \cite{Potts2019a}. We find that the intensity of the zero-bias peak monotonically decreases both with $B$-field (Fig. \ref{Fig4}(c)) and temperature (Fig. \ref{Fig4}(d)), and can therefore be interpreted as an underscreened spin-1 Kondo effect \cite{Pustilnik2001,Posazhennikova2005,Roch2009}. The temperature scaling can be described by Eq. \ref{eq_TK_temp}, and corresponds to $T_\mathrm{K}=780$\,mK ($G_0=0.07$\,$e^2$/$h$, $G_1=0.08$\,$e^2$/$h$) if using standard SU(2) parameters. \\

\section{Summary and Conclusion}

In summary, we have studied Kondo transport in parallel-coupled QDs in InAs nanowires. In the 1e regime we observe the spin-1/2 Kondo effect, the combined spin-orbital Kondo effect, and the orbital Kondo effect when the spin degeneracy is lifted. We demonstrate that finite inter-dot tunnel coupling inhibits the orbital Kondo effect by hybridizing the orbitals, and emphasize that the presence of a small energy gap can only be detected when the spin degeneracy is selectively lifted. The 2e regime exhibits a triplet ground state likely due to the formation of ring-like states, leading to a Kondo resonance at $B=0$. The possibility to isolate the different degeneracies makes this an ideal material system for studies of higher symmetry Kondo effects. \\

\begin{acknowledgments}
	The authors thank I-J. Chen and J. Paaske for fruitful discussions. This work was carried out with financial support from the Swedish Research Council (VR), NanoLund, the Knut and Alice Wallenberg Foundation (KAW), and the Crafoord Foundation. H.P. thankfully acknowledges funding from the Swiss National Science Foundation (SNSF) via Early PostDoc Mobility P2ELP2\_178221.
\end{acknowledgments}

\appendix
\section{Overview conductance measurement}

\begin{figure*}[ht]
	\includegraphics{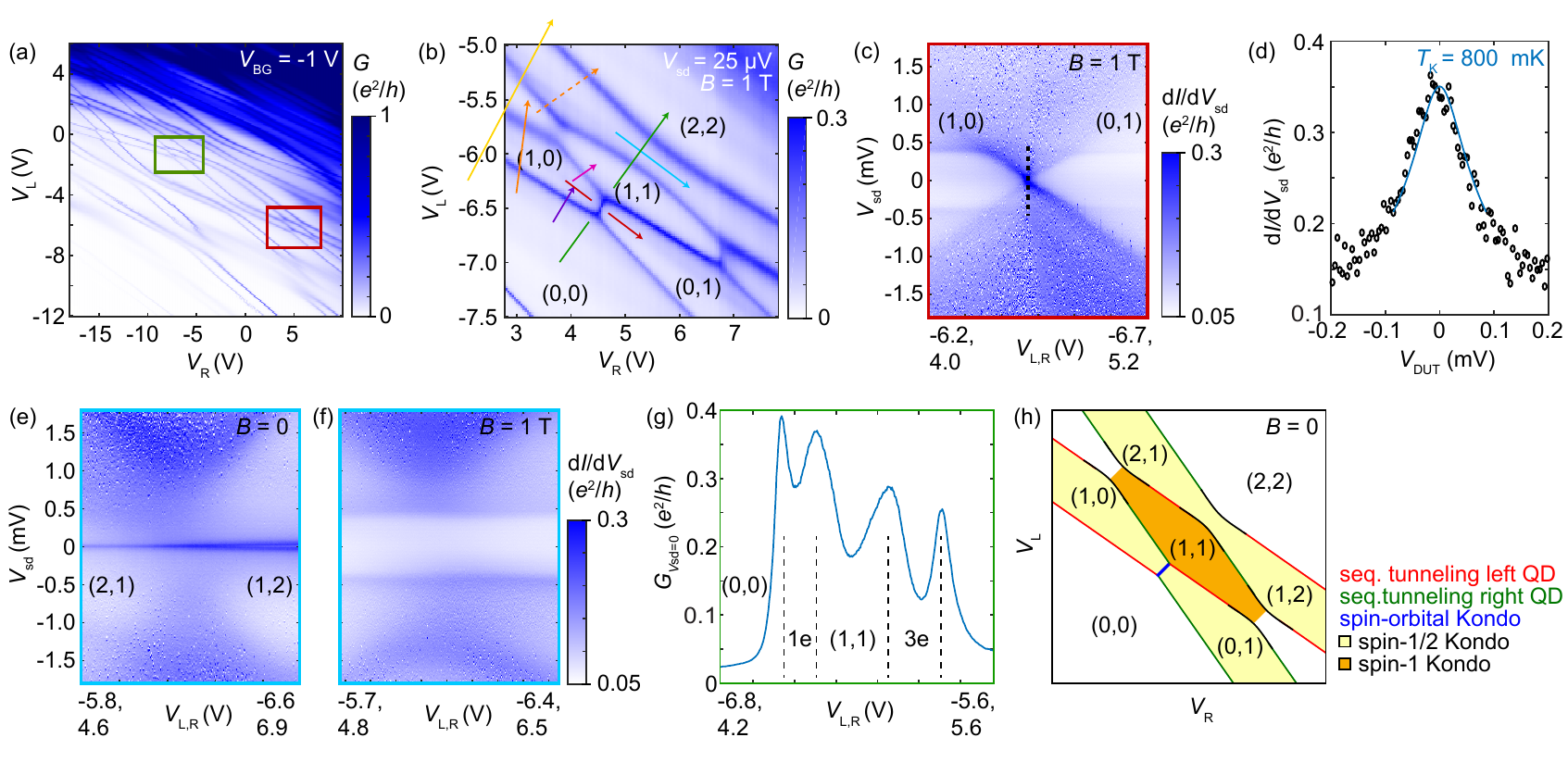}%
	\caption{(a) Conductance as a function of sidegate voltages. The red and green square indicate the orbital crossings which are studied. (b) Magnified conductance plot of the red orbital crossing, where relevant gate vectors are indicated. (c) Differential conductance versus source-drain voltage recorded along the red gate vector at $B=1$\,T. (d) Differential conductance as a function of voltage across the device measured at the sidegate voltages corresponding to the dashed line in (c). The data is fitted with Eq. \ref{eq:Kondo-Vds-fit} \cite{Pletyukhov2012}. (e)-(f) Transport recorded in the 3e regime along the cyan gate vector at $B=0$\,T, and $B=1$\,T, respectively. (g) Conductance $G$ at zero bias along the green gate vector. (h) Schematic representation of the orbital crossing and the different transport mechanisms at $B=0$\,T.  \label{FigSI1}}
\end{figure*}

Figure \ref{FigSI1}(a) shows conductance ($G$) for a wide range of sidegate voltages ($V_\mathrm{L}$,$V_\mathrm{R}$). The overview measurement shows that honeycombs with no discernible hybridization can be observed for specific orbital crossings. We explain this by the formation of ring-like states when the orbitals are coupled in two points, as investigated in detail in Ref. \cite{Potts2019a}. The orbital crossing showing the spin-orbital Kondo effect (c.f. Fig. \ref{Fig2} of the main article) is highlighted with a red square, while the additional crossing where the orbital Kondo effect is suppressed due to a small hybridization gap (c.f. Fig. \ref{Fig3} of the main article) is highlighted with a green square. We note that the data presented in Fig. \ref{Fig3} was obtained at a backgate voltage of $V_\mathrm{BG}=-0.8$\,V, while the overview here was obtained at $V_\mathrm{BG}=-1$\,V. In this supplementary material, we present additional measurements for the orbital crossing highlighted in red. The gate vectors which will be discussed in the following are indicated in the conductance measurement in Fig. \ref{FigSI1}(b) (same as Fig. \ref{Fig1}(b) of the main article).

\section{Extraction of $T_\mathrm{K}$ from the voltage dependence}

In the main article, the red gate vector where the orbitals are detuned from the (1,0) to the (0,1) regime is discussed in great detail, and the Kondo temperature ($T_\mathrm{K}$) is extracted from the temperature dependence of the zero-bias conductance peak. Alternatively, $T_\mathrm{K}$ can be determined from the source-drain voltage dependence of the zero-bias peak. Figure \ref{Fig1}(c) shows  differential conductance (d$I$/d$V_\mathrm{sd}$) versus the source-drain voltage ($V_\mathrm{sd}$) along the red gate vector at a magnetic field of $B=1$\,T (same as Fig. \ref{Fig2}(g) of the main article). Differential conductance as a function of applied voltage at the orbital degeneracy point (highlighted with a dashed line) is presented in Fig. \ref{FigSI1}(d). Note that the x-axis corresponds to the voltage applied across the device ($V_\mathrm{DUT}$), after subtracting the voltage drop across the series resistance of the amplifier and cryostat wiring and filters ($R_\mathrm{s}=16.5$\,k$\Omega$). For the spin-1/2 Kondo effect, Pletyukhov \textit{et. al.} have introduced that the voltage dependence of the differential conductance can be described by
\begin{equation}
\mathrm{d}I/\mathrm{d}V_\mathrm{sd}(V_\mathrm{sd}) = \mathrm{d}I/\mathrm{d}V_\mathrm{sd}(0)\left[1+\frac{\nu^{2}(2^{1/s1}-1)}{(1-b+b\nu^{s2})}\right]^{-s1}
\label{eq:Kondo-Vds-fit}
\end{equation}
with $s_1=0.32$, $b=0.05$, $s_2=1.26$ and $\nu=e(V_\mathrm{sd}-V_\mathrm{0})/(k_\mathrm{B}T_\mathrm{K}^*)$. The differential conductance at zero bias (d$I$/d$V_\mathrm{sd}(0)$), the Kondo temperature ($T_\mathrm{K}^*$), and $V_0$ are fitting parameters  \cite{Pletyukhov2012}. Using this equation to fit our experimental data for the orbital Kondo resonance, and after substituting $T_\mathrm{K}^*=1.8T_\mathrm{K}$, as suggested by Pletyukhov \textit{et. al.}\cite{Pletyukhov2012}, we extract a Kondo temperature of $T_\mathrm{K}=800$\,mK. We note that the Kondo temperature extracted from the voltage dependence is slightly higher compared to the one obtained by the temperature dependent data.

\section{Transport in the 3e regime}

In the main article, we showed that the orbital Kondo effect is absent if tunnel coupling between the two QDs leads to an avoided level crossing at $\Delta E_\mathrm{orb}=0$ (c.f. Fig. \ref{Fig3} of the main article). In order to illustrate that lifting the spin degeneracy can be key to detecting a small hybridization gap, we presented data from a different orbital crossing (highlighted with a green square in Fig. \ref{FigSI1}(a)). Here, we show transport measurements for the 3e regime of the orbital crossing highlighted with a red square in Fig. \ref{FigSI1}(a). Figures \ref{FigSI1}(e-f) show d$I$/d$V_\mathrm{sd}$ as a function of $V_\mathrm{sd}$ for $B=0$\,T and $B=1$\,T, when detuning the orbitals along the cyan gate vector. We observe a relatively large hybridization gap which can be seen both with and without magnetic field. This finding shows that adding electrons to the QD orbitals slightly alters the tunnel coupling between them. In the 1e regime the gate voltages were tuned in order to have the orbital Kondo effect (which requires the absence of a hybridization gap). In the 2e regime the two electrons strongly interact, resulting in two-electron states (c.f. Fig \ref{Fig4} of the main article). Finally, in the 3e regime the electrons also strongly interact, which we interpret as a gradual increase in tunnel coupling strength with particle number. Figure \ref{FigSI1}(g) shows the conductance at zero bias along the green gate vector (linecut at $V_\mathrm{sd} = 0$ in Fig. \ref{Fig2}(a) of the main article). The single electron peaks are clearly visible, indicating that the sample is not in the mixed valence regime for sidegate voltages in the center of the valleys \cite{Makarovski2007b}. This will be confirmed by estimating the tunnel coupling in Appendix \ref{App:Gamma}.

\begin{figure*}[ht]
	\includegraphics{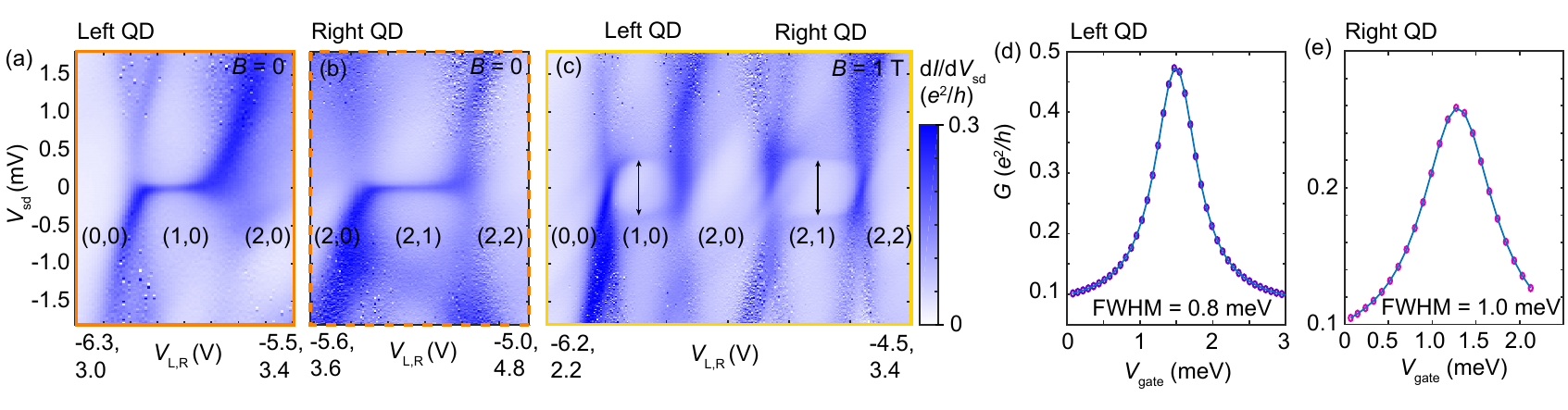}
	\caption{(a) Differential conductance versus source drain voltage as a function of gate vector through the unperturbed left and right QD levels (solid and dashed orange gate vector in Fig. \ref{FigSI1}(b)) for $B=0$\,T. (c) Corresponding measurement along the yellow gate vector for $B=1$\,T. (d)-(e) Conductance as a function of gate voltage for sequential tunneling through the left and right QD (purple and pink gate vector in Fig. \ref{FigSI1}(b)). The data is fitted using Eq. \ref{Eq:Gamma} to extract $\Gamma$. \label{FigSI2}}
\end{figure*}

Based on our observations, the different mechanisms leading to transport through this particular orbital crossing are summarized in Fig. \ref{FigSI1}(h): The red and green lines correspond to sequential tunneling through the left and right QD, respectively. The spin-1/2 Kondo effect is observed in the (0,1), (1,0), (1,2), and (2,1) regimes (yellow shading). In the 1e regime, the tunnel coupling between the two QDs is zero, leading to the spin-orbital Kondo effect at orbital degeneracy (blue line). In the (1,1) regime, a Kondo resonance is found due to a degenerate triplet ground state (orange shading). No orbital Kondo effect is found between the (2,0)-(1,1)-(0,2) regimes, explained by a small spin-orbit-induced mixing (see small avoided crossings in Fig. \ref{Fig4}(b)).

\section{Extraction of the leverarms and tunnel coupling}
\label{App:Gamma}

In Fig. \ref{FigSI2}(a,b) we present d$I$/d$V_\mathrm{sd}$ versus $V_\mathrm{sd}$ for gate vectors through the unperturbed levels of the left and right QD (corresponding to the solid/dashed orange vectors in Fig. \ref{FigSI1}(b)) at $B=0$\,T. A zero-bias peak can be observed due to the spin-1/2 Kondo effect in the (1,0) and (2,1) regimes, when the electron population on one of the QDs is odd. At $B=1$\,T (Fig. \ref{FigSI2}(c)) the spin degeneracy is lifted, resulting in a Zeeman gap. From the height of the Coulomb diamonds in Figs. \ref{FigSI2}(a)-(c), the charging energy $E_\mathrm{c}$ of the QDs can be extracted. The width of the Coulomb diamonds is then used to calculate the leverarms of the QDs with respect to each of the two sidegate voltages. We note that the Coulomb diamonds were recorded by sweeping both sidegates simultaneously (as indicated on the x-axis of Figs. \ref{FigSI2}(a)-(c)), and we used the slope of conductance lines in Fig. \ref{FigSI1}(b) in order to calculate an effective gate vector which depends only on one sidegate voltage. The relevant parameters for both QDs are summarized in Table \ref{Tab:leverarms}. Here, $\alpha_{V\mathrm{L/L}}$ denotes the leverarm of the left QD (L) with respect to the left sidegate voltage $V_\mathrm{L}$, and the other leverarms are labeled correspondingly.

\begin{table}[h]
	\begin{tabular}{|l|c|c|c|c|c|}
		\hline
		&$E_c$&$\alpha_{V\mathrm{L/L}}$&$\alpha_{V\mathrm{R/L}}$&$\alpha_{V\mathrm{L/R}}$&$\alpha_{V\mathrm{R/R}}$\\
		&(meV)&(meV/V)&(meV/V)&(meV/V)&(meV/V)\\
		\hline
		Left QD&6&14&4&-&-\\
		\hline
		Right QD&8&-&-&14&8\\
		\hline
	\end{tabular}
	\caption{Approximate values of the charging energy and the QD leverarms.}
	\label{Tab:leverarms}
\end{table}

Next, we extract the tunneling coupling ($\Gamma$) to the two QDs. This is relevant, since strictly speaking the empirical equation to calculate $T_\mathrm{K}$ from the temperature dependence is only valid as long as $\epsilon_0/\Gamma<-0.5$, where $\epsilon_0$ is the energy difference of the nearest lower energy state relative to the Fermi level of the contacts \cite{Goldhaber-Gordon1998}. 

$\Gamma$ can be estimated from the conductance due to sequential tunneling through a QD level. At low source-drain voltage and low temperature ($k_\mathrm{B}T \ll \Gamma$), the conductance $G$ as a function of gate voltage ($V_\mathrm{gate}$) can be described by a Lorentzian
\begin{equation}
G=\frac{G_\mathrm{max}}{1+(2\Delta V_\mathrm{gate}/\Gamma)^2}+c
\label{Eq:Gamma}
\end{equation}
where $G_\mathrm{max}$ is the peak conductance, $\Delta V_\mathrm{gate}=V_\mathrm{gate}-V_\mathrm{gate,0}$ is the detuning from the QD resonance centered at $V_\mathrm{gate,0}$, $\Gamma$ is the full width half maximum (FWHM) of the Lorentzian, and $c$ is a constant offset. The FWHM corresponds to the sum of the tunnel couplings to source and drain $\Gamma=\Gamma_\mathrm{S}+\Gamma_\mathrm{D}$. Figures \ref{FigSI2}(d,e) show linecuts through an orbital of the left and right QD, corresponding to the purple and pink gate vector in Fig. \ref{FigSI1}(b). We note that the gate vectors were chosen such that they are perpendicular to the conductance lines of the QD levels, after having converted the sidegate voltages into energies using the leverarms. Using equation \ref{Eq:Gamma}, $\Gamma\approx0.8$\,meV and $\Gamma\approx1$\,meV can be extracted for the left and right QD, respectively. We note that the sequential tunneling lines were measured at $V_\mathrm{sd}=25$\,$\mu$eV, leading to an additional broadening of the peak. The calculated $\Gamma$ therefore can be considered an upper bound. For the orbital Kondo effect, the measurement was done at orbital degeneracy in the 1e in the middle of the two charge state resonances (see Fig. \ref{Fig2}(a)). We therefore have $\epsilon_0=-0.5\cdot U_{1,2}=-1.5$\,meV, where $U_{1,2}=3$\,meV is the inter-dot Coulomb energy. With $\epsilon_0/\Gamma=-1.5<-0.5$ transport is in the Kondo regime, and the empirical equation to extract the Kondo temperature from the temperature dependence is therefore valid.

Finally, we extract the asymmetry $\Gamma_\mathrm{S}/\Gamma_\mathrm{D}$ for the two QDs based on the sequential tunneling conductance in Figs. \ref{FigSI2}(d)-(e). Using $G_\mathrm{max} = (4 \Gamma_\mathrm{S}\Gamma_\mathrm{D})/(\Gamma_\mathrm{S}+\Gamma_\mathrm{D})^2$\,$e^2$/$h$, we extract $\Gamma_\mathrm{S}/\Gamma_\mathrm{D}\approx20$ and $\Gamma_\mathrm{S}/\Gamma_\mathrm{D}\approx8$ for the left and right QD, respectively. We note that an estimation of the asymmetry based on $G_0$ in the Kondo regime would be less accurate since the electron temperature is comparable to $T_\mathrm{K}$ \cite{Pustilnik2004}. The asymmetry of the tunnel couplings of each QD also depends on the sidegate voltages due to a change of confinement potential.

\bibliography{DQD_Kondo}

\begin{thebibliography}{46}
\bibitem [{not(2019{\natexlab{b}})}]{note1}%
  \BibitemOpen
  \href@noop {} {} (\bibinfo {year} {2019}{\natexlab{b}}),\ \bibinfo {note} {we
  note that the small gap in the zero-bias peak in the 1e regime can be
  explained by the gate vector being not exactly aligned with the degeneracy of
  the QD levels.}\BibitemShut {Stop}%
\end{thebibliography}

\end{document}